\documentclass[traditabstract, longauth]{aa} 
\usepackage{graphicx}
\usepackage{txfonts}
\usepackage[breaklinks, colorlinks, citecolor=blue]{hyperref}
\usepackage{natbib}
\usepackage{Planck}
\newcommand{\Dang}{D_{\rm A}}

\newcommand{\Msun}{\rm M_\odot}

\newcommand{\thetafive}{\theta_{500}}

\newcommand{\sigf}{\sigma_{\thetafive}}

\newcommand{\Yscaled}{\tilde{Y}_{500}}
\newcommand{\sigscaled}{\tilde{\sigma}_{\thetafive}}
\newcommand{\YN}{Y_{20}}
\newcommand{\aN}{\alpha}

\begin{document}
\title{\textit{Planck} early results: Cluster Sunyaev-Zeldovich optical scaling relations}
\author{\small
Planck Collaboration:
N.~Aghanim\inst{45}
\and
M.~Arnaud\inst{55}
\and
M.~Ashdown\inst{53, 76}
\and
J.~Aumont\inst{45}
\and
C.~Baccigalupi\inst{67}
\and
A.~Balbi\inst{28}
\and
A.~J.~Banday\inst{75, 6, 60}
\and
R.~B.~Barreiro\inst{50}
\and
M.~Bartelmann\inst{73, 60}
\and
J.~G.~Bartlett\inst{3, 51}\thanks{Corresponding author: J.G. Bartlett, bartlett@apc.univ-paris7.fr}
% ...
}
\institute{\small
Aalto University Mets\"{a}hovi Radio Observatory, Mets\"{a}hovintie 114, FIN-02540 Kylm\"{a}l\"{a}, Finland\\
\and
Agenzia Spaziale Italiana Science Data Center, c/o ESRIN, via Galileo Galilei, Frascati, Italy\\
\and
Astroparticule et Cosmologie, CNRS (UMR7164), Universit\'{e} Denis Diderot Paris 7, B\^{a}timent Condorcet, 10 rue A. Domon et L\'{e}onie Duquet, Paris, France\\
\and
Atacama Large Millimeter/submillimeter Array, ALMA Santiago Central Offices Alonso de Cordova 3107, Vitacura, Casilla 763 0355, Santiago, Chile\\
\and
CITA, University of Toronto, 60 St. George St., Toronto, ON M5S 3H8, Canada\\
\and
CNRS, IRAP, 9 Av. colonel Roche, BP 44346, F-31028 Toulouse cedex 4, France\\
\and
California Institute of Technology, Pasadena, California, U.S.A.\\
\and
Centre of Mathematics for Applications, University of Oslo, Blindern, Oslo, Norway\\
\and
Centro de Astrof\'{\i}sica, Universidade do Porto, Rua das Estrelas, 4150-762 Porto, Portugal\\
\and
DAMTP, Centre for Mathematical Sciences, Wilberforce Road, Cambridge CB3 0WA, U.K.\\
\and
DSM/Irfu/SPP, CEA-Saclay, F-91191 Gif-sur-Yvette Cedex, France\\
\and
DTU Space, National Space Institute, Juliane Mariesvej 30, Copenhagen, Denmark\\
\and
Departamento de F\'{\i}sica, Universidad de Oviedo, Avda. Calvo Sotelo s/n, Oviedo, Spain\\
\and
Department of Astronomy and Astrophysics, University of Toronto, 50 Saint George Street, Toronto, Ontario, Canada\\
\and
Department of Physics \& Astronomy, University of British Columbia, 6224 Agricultural Road, Vancouver, British Columbia, Canada\\
\and
Department of Physics and Astronomy, University of Southern California, Los Angeles, California, U.S.A.\\
\and
Department of Physics, Gustaf H\"{a}llstr\"{o}min katu 2a, University of Helsinki, Helsinki, Finland\\
\and
Department of Physics, Princeton University, Princeton, New Jersey, U.S.A.\\
\and
Department of Physics, Purdue University, 525 Northwestern Avenue, West Lafayette, Indiana, U.S.A.\\
\and
Department of Physics, University of California, Berkeley, California, U.S.A.\\
\and
Department of Physics, University of California, One Shields Avenue, Davis, California, U.S.A.\\
\and
Department of Physics, University of California, Santa Barbara, California, U.S.A.\\
\and
Department of Physics, University of Illinois at Urbana-Champaign, 1110 West Green Street, Urbana, Illinois, U.S.A.\\
\and
Dipartimento di Fisica G. Galilei, Universit\`{a} degli Studi di Padova, via Marzolo 8, 35131 Padova, Italy\\
\and
Dipartimento di Fisica, Universit\`{a} La Sapienza, P. le A. Moro 2, Roma, Italy\\
\and
Dipartimento di Fisica, Universit\`{a} degli Studi di Milano, Via Celoria, 16, Milano, Italy\\
\and
Dipartimento di Fisica, Universit\`{a} degli Studi di Trieste, via A. Valerio 2, Trieste, Italy\\
\and
Dipartimento di Fisica, Universit\`{a} di Roma Tor Vergata, Via della Ricerca Scientifica, 1, Roma, Italy\\
\and
Discovery Center, Niels Bohr Institute, Blegdamsvej 17, Copenhagen, Denmark\\
\and
Dpto. Astrof\'{i}sica, Universidad de La Laguna (ULL), E-38206 La Laguna, Tenerife, Spain\\
\and
European Southern Observatory, ESO Vitacura, Alonso de Cordova 3107, Vitacura, Casilla 19001, Santiago, Chile\\
\and
European Space Agency, ESAC, Planck Science Office, Camino bajo del Castillo, s/n, Urbanizaci\'{o}n Villafranca del Castillo, Villanueva de la Ca\~{n}ada, Madrid, Spain\\
\and
European Space Agency, ESTEC, Keplerlaan 1, 2201 AZ Noordwijk, The Netherlands\\
\and
GEPI, Observatoire de Paris, Section de Meudon, 5 Place J. Janssen, 92195 Meudon Cedex, France\\
\and
Helsinki Institute of Physics, Gustaf H\"{a}llstr\"{o}min katu 2, University of Helsinki, Helsinki, Finland\\
\and
INAF - Osservatorio Astronomico di Padova, Vicolo dell'Osservatorio 5, Padova, Italy\\
\and
INAF - Osservatorio Astronomico di Roma, via di Frascati 33, Monte Porzio Catone, Italy\\
\and
INAF - Osservatorio Astronomico di Trieste, Via G.B. Tiepolo 11, Trieste, Italy\\
\and
INAF/IASF Bologna, Via Gobetti 101, Bologna, Italy\\
\and
INAF/IASF Milano, Via E. Bassini 15, Milano, Italy\\
\and
INRIA, Laboratoire de Recherche en Informatique, Universit\'{e} Paris-Sud 11, B\^{a}timent 490, 91405 Orsay Cedex, France\\
\and
IPAG: Institut de Plan\'{e}tologie et d'Astrophysique de Grenoble, Universit\'{e} Joseph Fourier, Grenoble 1 / CNRS-INSU, UMR 5274, Grenoble, F-38041, France\\
\and
Imperial College London, Astrophysics group, Blackett Laboratory, Prince Consort Road, London, SW7 2AZ, U.K.\\
\and
Infrared Processing and Analysis Center, California Institute of Technology, Pasadena, CA 91125, U.S.A.\\
\and
Institut d'Astrophysique Spatiale, CNRS (UMR8617) Universit\'{e} Paris-Sud 11, B\^{a}timent 121, Orsay, France\\
\and
Institut d'Astrophysique de Paris, CNRS UMR7095, Universit\'{e} Pierre \& Marie Curie, 98 bis boulevard Arago, Paris, France\\
\and
Institute of Astronomy and Astrophysics, Academia Sinica, Taipei, Taiwan\\
\and
Institute of Theoretical Astrophysics, University of Oslo, Blindern, Oslo, Norway\\
\and
Instituto de Astrof\'{\i}sica de Canarias, C/V\'{\i}a L\'{a}ctea s/n, La Laguna, Tenerife, Spain\\
\and
Instituto de F\'{\i}sica de Cantabria (CSIC-Universidad de Cantabria), Avda. de los Castros s/n, Santander, Spain\\
\and
Jet Propulsion Laboratory, California Institute of Technology, 4800 Oak Grove Drive, Pasadena, California, U.S.A.\\
\and
Jodrell Bank Centre for Astrophysics, Alan Turing Building, School of Physics and Astronomy, The University of Manchester, Oxford Road, Manchester, M13 9PL, U.K.\\
\and
Kavli Institute for Cosmology Cambridge, Madingley Road, Cambridge, CB3 0HA, U.K.\\
\and
LERMA, CNRS, Observatoire de Paris, 61 Avenue de l'Observatoire, Paris, France\\
\and
Laboratoire AIM, IRFU/Service d'Astrophysique - CEA/DSM - CNRS - Universit\'{e} Paris Diderot, B\^{a}t. 709, CEA-Saclay, F-91191 Gif-sur-Yvette Cedex, France\\
\and
Laboratoire Traitement et Communication de l'Information, CNRS (UMR 5141) and T\'{e}l\'{e}com ParisTech, 46 rue Barrault F-75634 Paris Cedex 13, France\\
\and
Laboratoire de Physique Subatomique et de Cosmologie, CNRS, Universit\'{e} Joseph Fourier Grenoble I, 53 rue des Martyrs, Grenoble, France\\
\and
Laboratoire de l'Acc\'{e}l\'{e}rateur Lin\'{e}aire, Universit\'{e} Paris-Sud 11, CNRS/IN2P3, Orsay, France\\
\and
Lawrence Berkeley National Laboratory, Berkeley, California, U.S.A.\\
\and
Max-Planck-Institut f\"{u}r Astrophysik, Karl-Schwarzschild-Str. 1, 85741 Garching, Germany\\
\and
Max-Planck-Institut f\"{u}r Extraterrestrische Physik, Giessenbachstra{\ss}e, 85748 Garching, Germany\\
\and
MilliLab, VTT Technical Research Centre of Finland, Tietotie 3, Espoo, Finland\\
\and
National University of Ireland, Department of Experimental Physics, Maynooth, Co. Kildare, Ireland\\
\and
Niels Bohr Institute, Blegdamsvej 17, Copenhagen, Denmark\\
\and
Observational Cosmology, Mail Stop 367-17, California Institute of Technology, Pasadena, CA, 91125, U.S.A.\\
\and
Optical Science Laboratory, University College London, Gower Street, London, U.K.\\
\and
SISSA, Astrophysics Sector, via Bonomea 265, 34136, Trieste, Italy\\
\and
SUPA, Institute for Astronomy, University of Edinburgh, Royal Observatory, Blackford Hill, Edinburgh EH9 3HJ, U.K.\\
\and
School of Physics and Astronomy, Cardiff University, Queens Buildings, The Parade, Cardiff, CF24 3AA, U.K.\\
\and
Space Research Institute (IKI), Russian Academy of Sciences, Profsoyuznaya Str, 84/32, Moscow, 117997, Russia\\
\and
Space Sciences Laboratory, University of California, Berkeley, California, U.S.A.\\
\and
Stanford University, Dept of Physics, Varian Physics Bldg, 382 Via Pueblo Mall, Stanford, California, U.S.A.\\
\and
Universit\"{a}t Heidelberg, Institut f\"{u}r Theoretische Astrophysik, Albert-\"{U}berle-Str. 2, 69120, Heidelberg, Germany\\
\and
Universit\'{e} Denis Diderot (Paris 7), 75205 Paris Cedex 13, France\\
\and
Universit\'{e} de Toulouse, UPS-OMP, IRAP, F-31028 Toulouse cedex 4, France\\
\and
University of Cambridge, Cavendish Laboratory, Astrophysics group, J J Thomson Avenue, Cambridge, U.K.\\
% ...
}

\abstract{We present the Sunyaev-Zeldovich (SZ) signal-to-richness scaling relation ($Y_{500}-N_{200}$) for the MaxBCG cluster catalogue.  Employing a multi-frequency matched filter on the \Planck\ sky maps, we measure the SZ signal for each cluster by adapting the filter according to weak-lensing calibrated mass-richness relations ($N_{200}-M_{500}$).  We bin our individual measurements and detect the SZ signal down to the lowest richness systems ($N_{200}=10$) with high significance, achieving a detection of the SZ signal in systems with mass as low as $M_{500}\approx 5\times 10^{13}\,\Msun$.   

The observed $Y_{500}-N_{200}$ relation is well modeled by a power law over the full richness range.  It has a lower normalisation at given $N_{200}$ than predicted based on X-ray models and published mass-richness relations.  An X-ray subsample, however, does conform to the predicted scaling, and model predictions do reproduce the relation between our measured bin-average SZ signal and measured bin-average X-ray luminosities.  At fixed richness, we find an intrinsic dispersion in the  $Y_{500}-N_{200}$  relation of 60\% rising to of order 100\% at low richness.  

Thanks to its all-sky coverage, \Planck\ provides observations for more than 13,000 MaxBCG clusters and an unprecedented SZ/optical data set, extending the list of known cluster scaling laws to include SZ-optical properties.  The data set offers essential clues for models of galaxy formation.  Moreover, the lower normalisation of the SZ-mass relation implied by the observed  SZ-richness scaling has important consequences for cluster physics and cosmological studies with SZ clusters.
}

\authorrunning{Planck Collaboration}
\titlerunning{SZ- optical richness relation}
   \maketitle

\section{Introduction}

Galaxy cluster properties follow simple scaling laws \citep[see e.g.][for recent reviews]{rosati2002, voit2005}.  This attests to a remarkable consistency in the cluster population and motivates the use of clusters as cosmological probes.  These scaling laws also provide important clues to cluster formation, and relations involving optical properties, in particular, help uncover the processes driving galaxy evolution.  

\begin{figure}
\centering
\includegraphics[width=9cm]{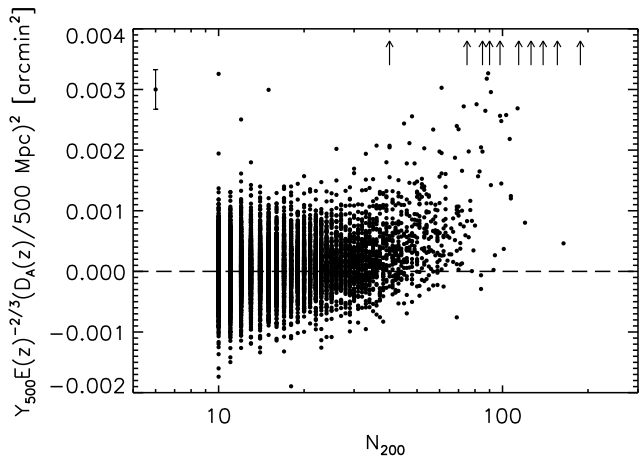}
\caption{Individual scaled SZ signal measurements, $\Yscaled$, for the MaxBCG catalogue as a function of richness $N_{200}$.  We do not plot individual error bars to avoid saturating the figure.  The error bar drawn in the upper left represents the {\em median} uncertainty over the entire population; in general, the uncertainty increases towards low richness.  The SZ signal measurements are expressed as the Compton $y$ parameter integrated over a sphere out to $R_{500}$, scaled in redshift according to the self-similar model and placed at a fiducial angular distance of $500\,$Mpc.   Each point represents the result of the matched filter applied to an individual cluster in the catalogue.  Upward pointing arrows indicate values beyond the plotted range.  The radius $R_{500}$, and hence the filter size, is set from the mass of each cluster determined via the weak-lensing calibrated $M_{500}-N_{200}$ relation given by \citet{johnston2007}.  The results are nearly
the same for the relation given by \citet{rozo2009}}
\label{fig:Yall}
\end{figure}

The Sunyaev-Zeldovich (SZ) effect \citep{sz1972, birkinshaw1999} opens a fresh perspective on cluster scaling laws, and the advent of large-area SZ surveys furnishes us with a powerful new tool \citep{carlstrom2002}.  Proportional to ICM mass and temperature, the thermal SZ effect probes the gas in a manner complementary to X-ray measurements, 
giving a more direct view of the gas mass and energy content.   Ground-based instruments, such as the Atacama Cosmology Telescope [(ACT), \citep{swetz2008}], the South Pole Telescope [SPT, \citep{carlstrom2009}] and APEX-SZ \citep{dobbs2006}, are harvesting a substantial crop of scientific results.  They are producing, for the first time, SZ-selected catalogues and using them to constrain cosmological parameters \citep{staniszewski2009, marriage2010, sehgal2010, vanderlinde2010}.  

The \Planck\footnote{\Planck\ (http://www.esa.int/\Planck ) is a project of the European Space
Agency (ESA) with instruments provided by two scientific consortia funded by ESA member
states (in particular the lead countries France and Italy), with contributions from NASA
(USA) and telescope reflectors provided by a collaboration between ESA and a scientific
consortium led and funded by Denmark.}\ consortium has published its first scientific results \citep{planck2011-1.1} and released the \Planck\ Early Release Compact Source Catalogue (ERCSC) \citep{planck2011-1.10}, which includes the \Planck\ Early SZ (ESZ) all-sky cluster list \citep{planck2011-5.1a}.  \Planck\ (\cite{tauber2010a}; \cite{planck2011-1.1}) is the third generation space mission to measure the anisotropy of the cosmic microwave background (CMB).  It observes the sky in nine frequency bands covering 30--857\,GHz with high sensitivity and angular resolution from 31\arcm--5\arcm.  The Low Frequency Instrument (LFI; \cite{Mandolesi2010}; \cite{Bersanelli2010}; \cite{planck2011-1.4}) covers the 30, 44, and 70\,GHz bands with amplifiers cooled to 20\,\hbox{K}.  The High Frequency Instrument (HFI; \cite{Lamarre2010}; \cite{planck2011-1.5}) covers the 100, 143, 217, 353, 545, and 857\,GHz bands with bolometers cooled to 0.1\,\hbox{K}.  Polarization is measured in all but the highest two bands (\cite{Leahy2010}; \cite{Rosset2010}).  A combination of radiative cooling and three mechanical coolers produces the temperatures needed for the detectors and optics (\cite{planck2011-1.3}).  Two Data Processing Centres (DPCs) check and calibrate the data and make maps of the sky (\cite{planck2011-1.7}; \cite{planck2011-1.6}).  \Planck's sensitivity, angular resolution, and frequency coverage make it a powerful instrument for galactic and extragalactic astrophysics, as well as cosmology.  Early astrophysics results are given in \citet{planck2011-5.1a}-\citet{planck2011-7.13}.

\Planck\ early results on clusters of galaxies are presented in this {\em paper\/} and in \citep{planck2011-5.1a,planck2011-5.1b,planck2011-5.2a,planck2011-5.2b}.  In the present work, we use \Planck\ SZ measurements at the locations of MaxBCG clusters \citep{koester2007b} to extract the SZ signal-richness scaling relation.  There are several optical cluster catalogs \citep{wen2009, hao2010, szabo2010} available from the Sloan Digital Sky Survey \citep[SDSS]{york2000}.  For this initial study, we chose the MaxBCG catalogue for its large sample size, wide mass range and well-characterized selection function, and because its properties have been extensively studied.  In particular, we benefit from weak-lensing mass measurements and mass-richness relations \citep{johnston2007, mandelbaum2008a, sheldon2009, rozo2009}.
A combined SZ-optical study over such a large catalogue is unprecedented and \Planck\ is a unique SZ instrument for this task, as its all-sky coverage encompasses the complete SDSS area and the full MaxBCG cluster sample.  

Our analysis methodology follows that of the accompanying paper on the SZ properties of X-ray  selected clusters \citep{planck2011-5.2a}.  Although the individual SZ measurements in both cases generally have low signal-to-noise, we extract the statistical properties of the ICM --- mean relations and their dispersion --- by averaging over the large sample.  The approach enables us to study the properties of a much larger and representative sample of clusters than otherwise possible.

\begin{figure*}
\centering
\includegraphics[scale=0.5]{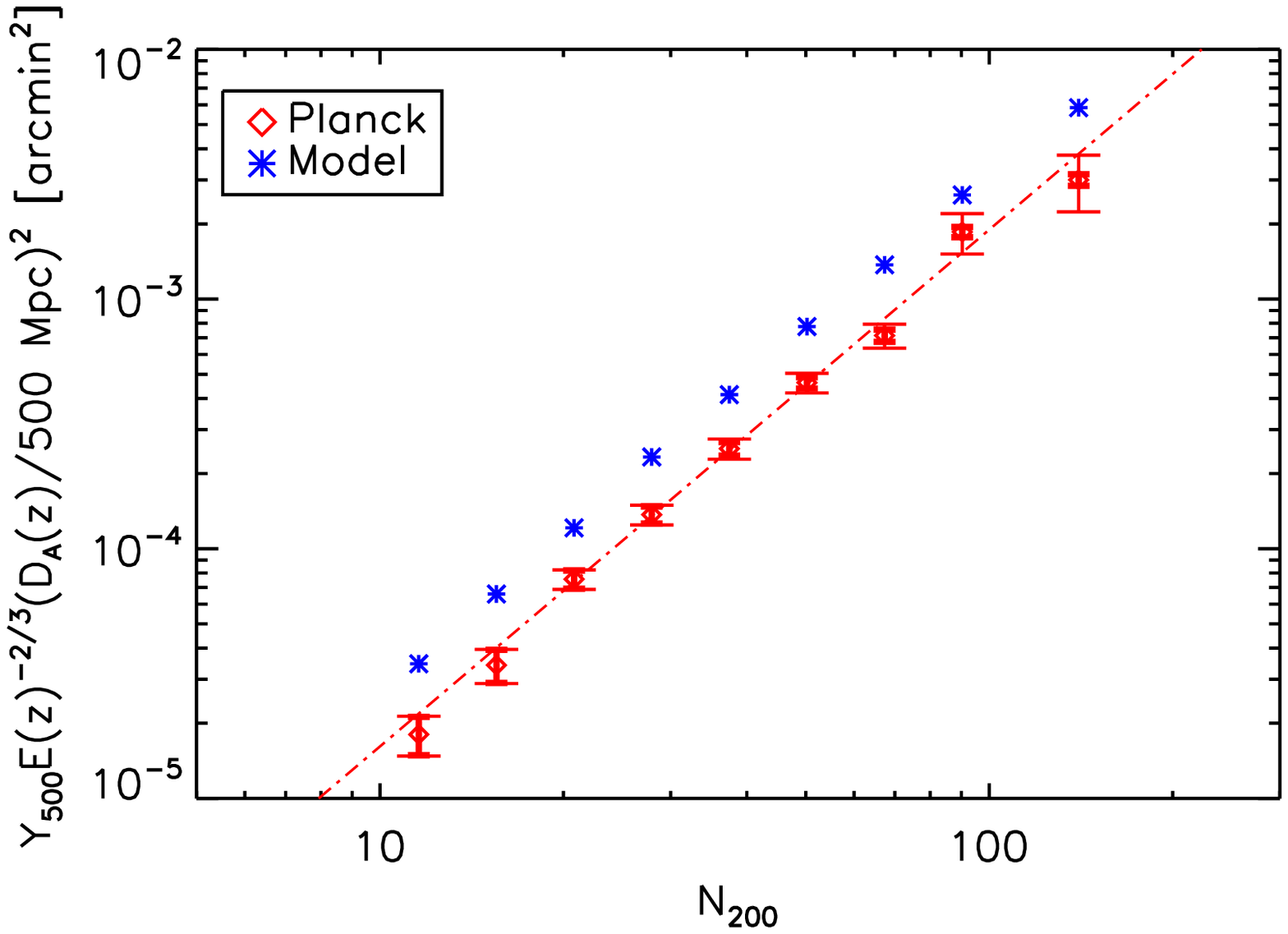}
\includegraphics[scale=0.5]{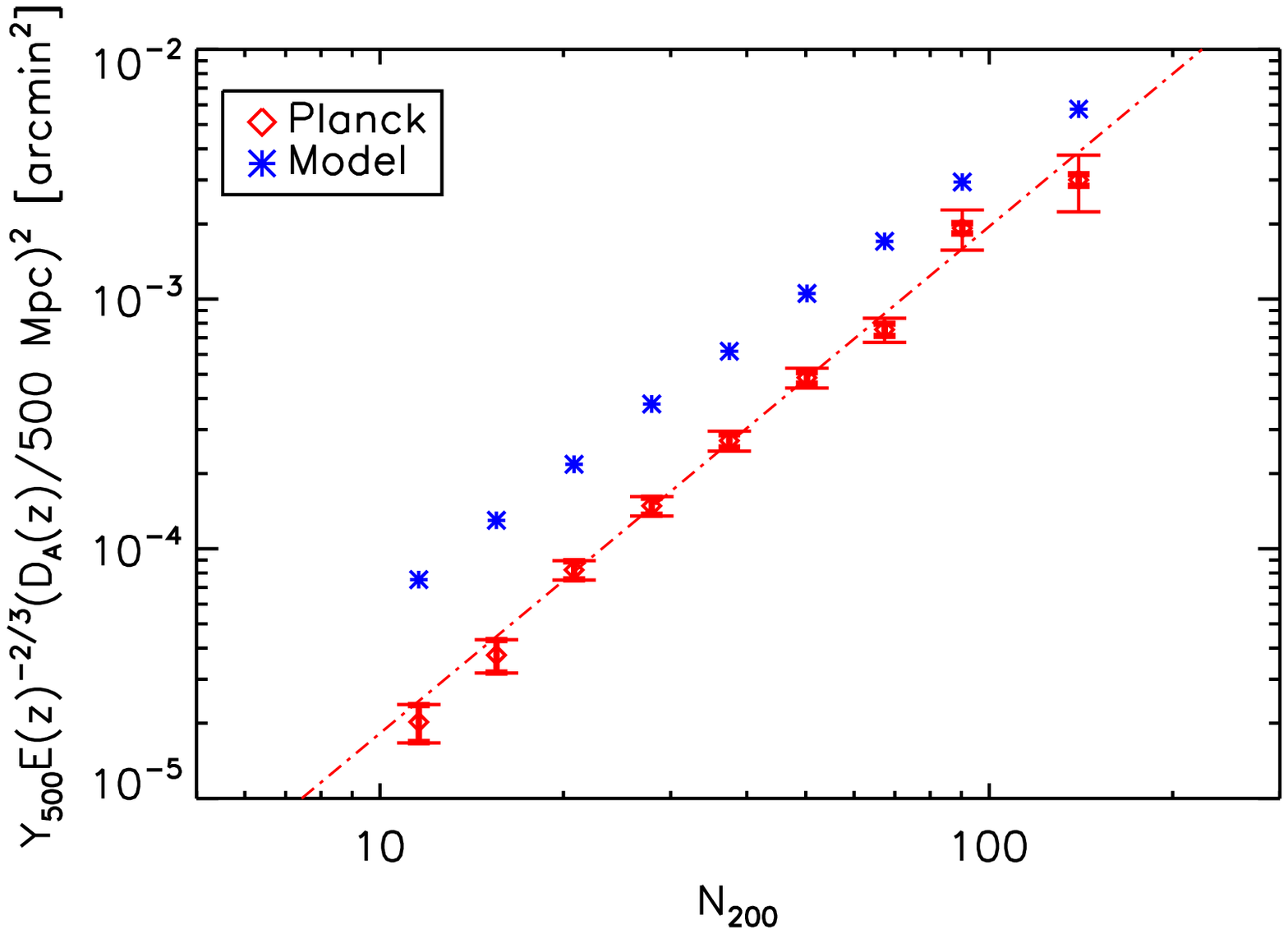}
\caption{Scaled SZ signal measurements, $\Yscaled$, binned by richness, $N_{200}$.  The left-hand panel presents the results for the \citet{johnston2007} $M_{500}-N_{200}$ relation, the right-hand panel for the \citet{rozo2009} relation.  In each case, the red diamonds show the bin-average, redshift-scaled $\Yscaled$ calculated as the weighted mean of all individual measurements (e.g., Fig. \ref{fig:Yall}) in the bin, where the weights are taken from the estimated filter noise.  The thick error bars show the corresponding uncertainty on the bin-average SZ signal, while the lighter error bars indicate the uncertainty found by bootstrap analysis;  they are larger due to the presence of intrinsic scatter within the bins, most notable at high richness (see Fig. \ref{fig:scatter}).  The blue points represent the model prediction for each bin found by averaging, with the same weights as the data, the SZ signal expected from the $Y_{500}-M_{500}$ \citep[STD case]{arnaud2010} and corresponding $M_{500}-N_{200}$ relations.  The \Planck\ measurements are little affected by choice of mass-richness relation, while the model points move significantly upward with the \citet{rozo2009} mass calibration.  Dashed lines in both panels show the best fit power-law to the \Planck\ individual cluster data points (i.e., prior to binning, as shown in Fig. \ref{fig:Yall}); the parameters for these fits are given in Table \ref{tab:fits}.}
\label{fig:Ybinned}
\end{figure*}

The SZ-richness relation adds a new entry to the complement of cluster scaling laws and additional constraints on cluster and galaxy evolution models.  With a mass-richness relation, we can also derive the SZ signal-mass relation.  This is a central element in predictions for the diffuse SZ power spectrum and SZ cluster counts.  Poor knowledge of the relation represents an important source of modeling uncertainty.  Low mass systems, for example, contribute a large fraction of the SZ power, but we know very little about their SZ signal. 

We organise the {\em paper\/} as follows: the next section presents the data used, both the \Planck\ maps and the MaxBCG catalogue and pertinent characteristics.  Section 3 details our SZ measurements based on a multi-frequency matched filter, and outlines some of the systematic checks.  In Sect.~4 we present our basic results and in 
Sect.~5 compare them to model expectations.  Section 6 concludes.

\subsection{Conventions and Notation}
\label{sec:notation}
In the following, we adopt a flat fiducial cosmology with \mbox{$\Omega_{\rm M}=0.3$} with the remainder of the critical density made up by a cosmological constant.  We express the Hubble parameter at redshift $z$ as \mbox{$H(z)=H_0E(z)=(h\times 100\,\kmsMpc) E(z)$} with \mbox{$h=0.7$}.  Cluster radii are expressed in terms of $R_\Delta$, the radius inside of which the mean mass overdensity equals \mbox{$\Delta\times \rho_{\rm c}(z)$}, where \mbox{$\rho_{\rm c}(z)=3H^2(z)/8\pi G$} is the critical density at redshift $z$.  Similarly, we quote masses as \mbox{$M_\Delta=\Delta (4\pi/3)R^3_\Delta\rho_{\rm c}$}.  
We note that, in contrast, optical cluster studies, and in particular the MaxBCG group, frequently employ radii and masses scaled
to the mean matter density, rather than the critical density.  For example, it is standard practice to refer to quantities measured within $R_{200{\rm b}}$, where the overdensity of 200 is defined with respect to the background density (this corresponds to $R_{60}$ at $z=0$ and $R_{155}$ at $z=1$).  For richness we will use the MaxBCG $N_{200}$, defined as the number of red galaxies with $L>0.4L_*$ within $R_{200{\rm b}}$.  Richness $N_{200}$ is the only quantity in this work defined relative to the mean background density.

We characterize the SZ signal with the Compton-$y$ parameter integrated over a sphere of radius $R_{500}$ and expressed in arcmin$^2$: \mbox{$Y_{500}= (\sigma_{\rm T}/m_{\rm e} c^2)\int_0^{R_{500}}\; P dV/\Dang^2(z)$}, where $\Dang$ denotes angular distance, $\sigma_{\rm T}$ is the Thomson cross-section, $c$ the speed of light,  $m_{\rm e}$ the electron rest mass and $P=n_{\rm e} kT$  is the pressure, defined as the product of the electron number density and temperature, $k$ being the Boltzmann constant.
The use of this spherical, rather than cylindrical, quantity is possible because we adopt a template SZ profile when using the matched filter (discussed below).  We bring our measurements to $z=0$ and a fiducial angular distance assuming self-similar scaling in redshift.  To this end, we introduce the intrinsic cluster quantity (an ``absolute SZ signal strength'') \mbox{$\Yscaled\equiv Y_{500} E^{-2/3}(z)(\Dang(z)/500\, {\rm Mpc})^2$}, also expressed in arcmin$^2$.

\section{Data Sets}

We base our study on \Planck\ SZ measurements at the positions of clusters in the published MaxBCG cluster catalogue.  

\subsection{The MaxBCG Optical Cluster Catalogue} 
\label{sec:maxbcg}

The MaxBCG catalogue \citep{koester2007a, koester2007b} is derived from Data Release~5 (DR5) of the Sloan Digital Sky Survey \citep{york2000},  covering an area of $7500\,$~deg$^2$ in the Northern hemisphere.  Galaxy cluster candidates were extracted by color, magnitude and a spatial filter centered on galaxies identified as the Brightest Cluster Galaxy (BCG). The catalogue provides position, redshift, richness and total luminosity for each candidate.  In the following we will only use the richness $N_{200}$, defined as the number of red-sequence galaxies with $L>0.4L_*$ and within a projected radius at which the cluster interior mean density equals 200 times the {\em mean} background density at the redshift of the cluster (see \citet{koester2007b} for details and the remark in Section \ref{sec:notation}).
The catalogue consists of 13,823 galaxy clusters over the redshift range $0.1<z<0.3$, with $90\%$ purity and $85\%$ completeness for $10<N_{200}<190$ as determined from simulations.

A valuable characteristic for our study is the wide mass range spanned by the catalogue.  Another is the fact that numerous authors have studied the catalogue, providing extensive information on its properties.  In particular, \citet{sheldon2009} and \citet{mandelbaum2008a} have published mass estimates from weak gravitational lensing analyses, which \citet{johnston2007} and \citet{rozo2009} use to construct mass-richness ($M_{500}-N_{200}$) relations.  We apply this relation, as outlined below, to adapt our SZ filter measurements for each individual cluster according to its given richness, $N_{200}$, as well as in our model predictions.

In their discussion, \citet{rozo2009} identify the differences between the \citet{sheldon2009} and \citet{mandelbaum2008a} mass estimates and the impact on the deduced mass-richness relation.  They trace the systematically higher mass estimates of \citet{mandelbaum2008a} to these authors' more detailed treatment of photometric redshift uncertainties \citep{mandelbaum2008b}.  Moreover, they note that \citet{johnston2007}, when employing the \citet{sheldon2009} measurements, used an extended MaxBCG catalogue that includes objects with $N_{200}<10$, where the catalogue is known to be incomplete.  These two effects lead \citet{rozo2009} to propose a flatter mass-richness relation with higher normalisation than the original \citet{johnston2007} result.  In the following, we perform our analysis with both relations; specifically, using the fit in Table 10 for the $M_{500}-N_{200}$ relation of \cite{johnston2007}, and Eqs. (4), (A20) and (A21) of \citet{rozo2009}.

\subsection{\Planck\ Data}

We use the six HFI channel temperature maps (prior to CMB removal) provided by the DPC and whose characteristics are given in \citet{planck2011-1.7}. 
These maps correspond to the observations of intensity in the first ten months of survey by Planck, still allowing complete sky coverage.  Hence, they give us access to the entire SDSS survey area and complete MaxBCG catalogue.  After masking bad pixels and contaminated regions (e.g., areas where an individual frequency map has a point source at $>10\sigma$), we have \Planck\ observations for 13,104  of the 13,823 clusters in the MaxBCG catalogue.

\begin{figure*}
\centering
\includegraphics[scale=0.5]{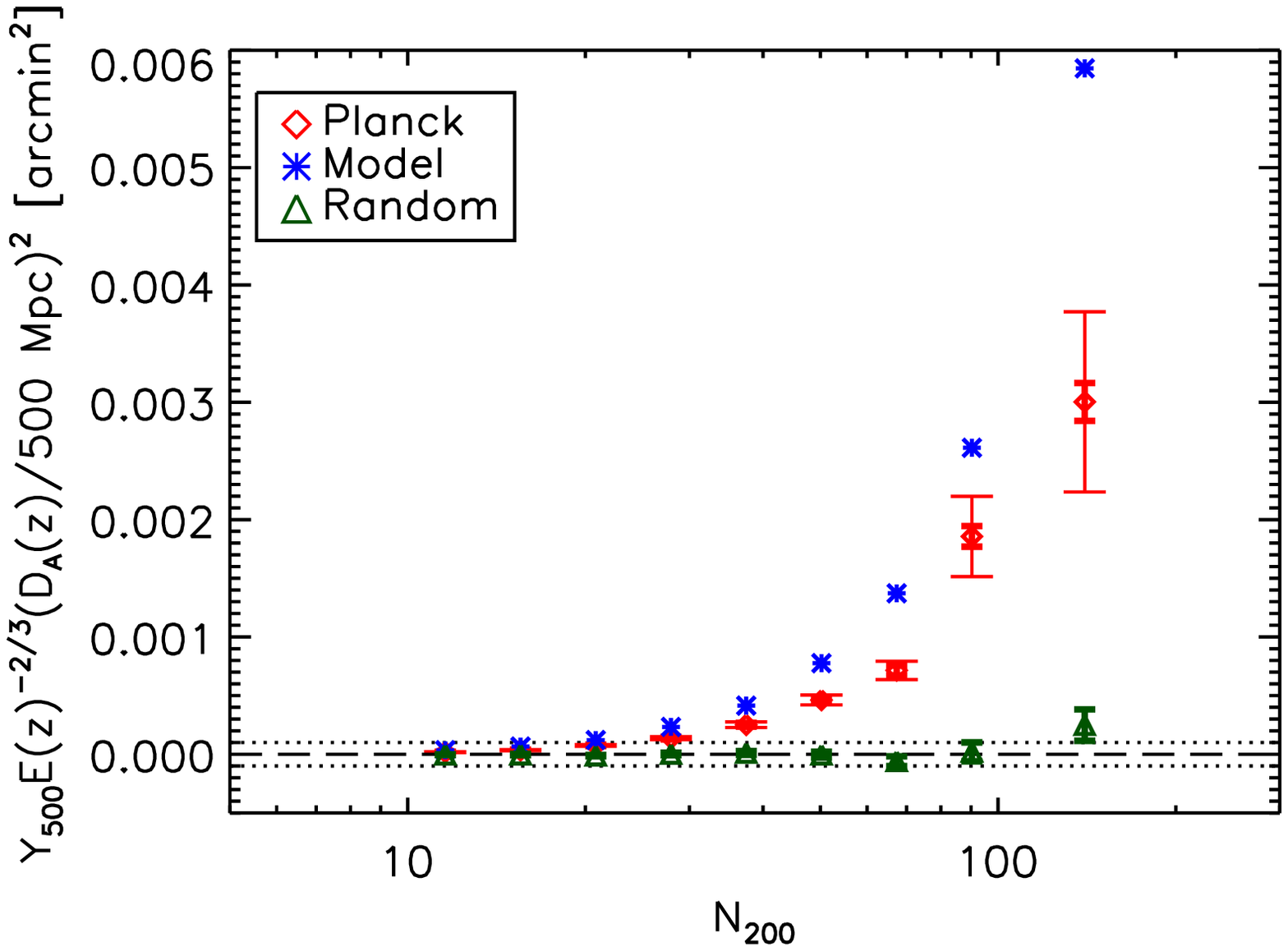}
\includegraphics[scale=0.5]{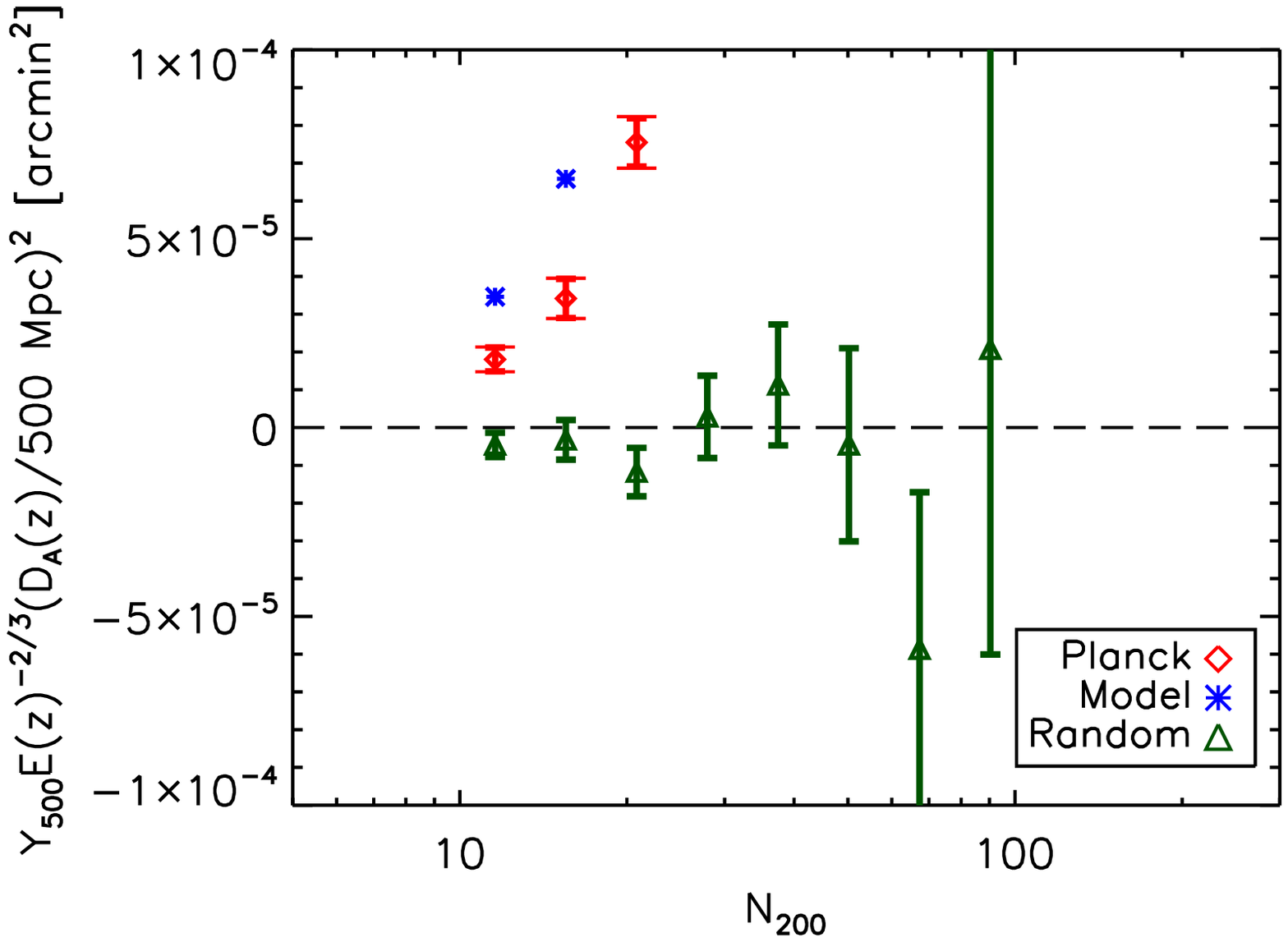}
\caption{Null test performed by randomising the angular positions of the clusters.  The red diamonds show the bin-average, redshift-scaled measurements, $\Yscaled$, as reported in the left-hand side of Figure~\ref{fig:Ybinned} with their corresponding measurement and bootstrap uncertainties; blue stars are the same model points.  The green triangles present the bin-averages for the randomised catalogue with uncertainties given only by the SZ measurement errors.  Results for the randomised catalogue are consistent with zero within their uncertainties.  By comparison, the values for the real catalogue represent highly significant detections of the SZ signal in all richness bins.  {\em Left-hand panel:\/} Results over the full richness range.  {\em Right-hand panel:\/} Zoom into the region indicated by the dotted lines in the left-hand panel to highlight the low-richness end.}
\label{fig:nulltest}
\end{figure*}

\section{SZ Measurements}

We extract the SZ signal at the position of each MaxBCG cluster by applying a multi-frequency matched filter \citep{herranz2002, melin2006} to the six \Planck\ temperature maps.  The technique maximises the signal-to-noise of objects having the known frequency dependence of the thermal SZ effect and the expected angular profile.  The filter returns the amplitude of the template, which we then convert into integrated SZ signal, $Y_{500}$, within $R_{500}$.  It also returns an estimate of the local noise through the filter, $\sigf$, due to instrumental noise and astrophysical emissions.  The same procedure is used in \citet{planck2011-5.2a}.  We refer the reader to \citet{melin2006, melin2010} for details.

\subsection{SZ Model Template}
For the filter's spatial template we adopt the empirical universal pressure profile of \citet{arnaud2010}, deduced from X-ray studies of the REXCESS cluster sample \citep{bohringer2007}:
\begin{equation}
P(r) \propto \frac{1}{x^\gamma (1+x^\alpha)^{(\beta-\gamma)/\alpha}}
\end{equation}
where the physical radius $r$ is scaled to $x=r/r_{\rm s}$, with $r_{\rm s}=R_{500}/c_{500}$.  For the standard self-similar case [ST case in Appendix B of \citet{arnaud2010}], $c_{500}=1.156$ and the exponents are $\alpha=1.0620$, $\beta=5.4807$, $\gamma=0.3292$.  The normalisation is arbitrary for purposes of the matched filter.  The SZ signal being proportional to the gas pressure, we find the filter template by integrating along the line-of-sight and expressing the result in terms of projected angles: $x=\theta/\theta_{\rm s}$.  We truncate the filter at $5\thetafive$,
containing more than 95\% of the signal for the model.

\subsection{Application of the Filter}

We apply the matched filter to each cluster in the MaxBCG catalogue, using the mass-richness relation, $M_{500}-N_{200}$, to define $R_{500}$ and set the angular scale $\thetafive=R_{500}/\Dang(z)$.  The filter effectively samples the cluster SZ signal along a  cone out to a transverse angular radius of $5\thetafive$, and returns the normalisation for the template.  We apply a geometric factor based on the template SZ profile to convert the deduced total SZ signal along the cone to an equivalent $Y_{500}$ value, the SZ signal integrated within a sphere of physical radius $R_{500}$.  To account for the redshift range of the catalogue, we scale these measurements according to self-similar expectations to redshift $z=0$ and a fiducial angular distance of $500\,$Mpc: $\Yscaled\equiv Y_{500}E^{-2/3}(z)(\Dang(z)/500\,{\rm Mpc})^2$.
We accordingly adapt the estimated filter noise $\sigf$ to uncertainty $\sigscaled$ on these scaled SZ signal measurements.  The results of this procedure when using the \citet{johnston2007} mass-richness relation are shown in Figure \ref{fig:Yall}.

\subsection{Systematic Effects}

As in the other four \Planck\ SZ papers \citep{planck2011-5.1a,planck2011-5.1b,planck2011-5.2a,planck2011-5.2b}, we have carried out various tests to ensure the robustness of the \Planck\ SZ measurements.  They included investigation of the cluster size-flux degeneracy, evaluation of the impact of the assumed pressure profile used for the \Planck\ cluster detection, of beam-shape effects, color corrections, potential contamination by point sources, as well as an overall error budget estimation.  We refer the reader to Sec.~6 of \citet{planck2011-5.1a} for an extensive description of this common analysis.

To complete this investigation in the present work, we repeated our entire analysis, changing both the instrument beams and adopted SZ profiles.  In the former instance, we varied the beams at all frequencies together to the extremes of their associated uncertainties as specified by the DPC \citep{planck2011-1.7}.  All beams were increased or all decreased in lock-step to maximize any effect.  To investigate the profile, we re-extracted the SZ signal using a non-standard SZ signal-mass scaling, and separately for cool-core and morphologically disturbed SZ profiles (based on the work of \citet{arnaud2010}).  In all cases, the impact on the measurements was of order a few percent
and thus negligibly impacts our conclusions.

\begin{table}[tmb] 
\begingroup 
\newdimen\tblskip \tblskip=5pt
\caption{Scaled Planck SZ signal measurements $\Yscaled$ binned by $N_{200}$ for the \cite{rozo2009} mass-richness relation.  Given $\Yscaled$ values are the measurement-noise weighted mean in the bin.  The statistical uncertainty corresponds to the measurement-noise uncertainty on the weighted mean, while the total uncertainty expresses the standard deviation of the weighted mean from an ensemble of bootstrap samples.  This table is plotted as the red diamonds and error bars in the right-hand panel of Figure \ref{fig:Ybinned}.}
\label{tab:binvalues}
\nointerlineskip
\vskip -3mm
\footnotesize 
\setbox\tablebox=\vbox{ %
\newdimen\digitwidth 
\setbox0=\hbox{\rm 0}
\digitwidth=\wd0
\catcode`*=\active
\def*{\kern\digitwidth}
\newdimen\signwidth
\setbox0=\hbox{+}
\signwidth=\wd0
\catcode`!=\active
\def!{\kern\signwidth}
\halign{ #\hfil \hspace{0.2cm} & \hfil #\hfil \hspace{0.2cm} & \hfil #\hfil \hspace{0.2cm} & \hfil #\hfil \cr
\noalign{\doubleline}
$N_{200}$ & $\Yscaled/(10^{-5}\,{\rm arcmin}^2)$ & Stat. Uncertainty & Total Uncertainty \cr
\noalign{\vskip 3pt\hrule\vskip 5pt}
10-13     & 2.0    & $\pm 0.3$    & $\pm 0.3$ \cr
14-17     & 3.8    & $\pm 0.6$    & $\pm 0.6$ \cr
18-24     & 8.2    & $\pm 0.7$    & $\pm 0.7$ \cr
25-32     & 15     & $\pm 1   $    & $\pm 1   $ \cr
33-43     & 27     & $\pm 2   $    & $\pm 2   $ \cr
44-58     & 48     & $\pm 3   $    & $\pm 4   $ \cr
59-77     & 76     & $\pm 4   $    & $\pm 8   $ \cr
78-104   & 190   & $\pm 9   $    & $\pm 40 $ \cr
$>$105   & 300   & $\pm 20 $    & $\pm 80 $ \cr
\noalign{\vskip 5pt\hrule\vskip 3pt}}}
\endPlancktable 
\endgroup
\end{table}

\section{Results}

Our basic measurements are the set of individual scaled SZ signal values $\Yscaled$ for each MaxBCG cluster, given as a function of richness $N_{200}$ in Figure~\ref{fig:Yall} for the \citet{johnston2007} mass calibration.  At high richness we can detect by eye a slight upturn of the points.  Except for the most massive objects, however, the signal-to-noise of the individual measurements is small, in most cases well below unity.  This is as expected given the masses of the clusters and the \Planck\ noise levels.

To extract the signal, we bin these $\Yscaled$ values by richness and calculate the bin averages as the noise-weighted mean of all individual $i=1,.., N_{\rm b}$ measurements falling within the bin: $\langle \Yscaled\rangle_{\rm b} = (\sum_i \Yscaled(i)/\sigscaled^2(i))/(\sum_i 1/\sigscaled^2(i))$.  We plot the result as the red diamonds in Figure~\ref{fig:Ybinned}.  The bold error bars represent only the statistical uncertainty associated with the SZ signal measurements: $\sigma^{-2}_{\rm b} = \sum_i 1/\sigscaled^2(i)$ (in some cases the error bars are hidden by the size of the data point in the figure).  The left-hand panel of the figure shows results using the \cite{johnston2007} mass calibration, while the right-hand side gives results for the \cite{rozo2009} mass calibration.  The individual SZ signal measurements are not sensitive to this choice: the different calibrations do modify the adopted filter size, but the
impact on the measured signal is small.  

We quantify the significance of the SZ detection using a null test: we perform an identical analysis on the MaxBCG catalogue after first randomising the cluster angular positions within the SDSS DR5 footprint.  In this analysis we are therefore attempting to measure SZ signal with the same set of filters, but now positioned randomly within the SDSS survey.  The result is shown in Figure~\ref{fig:nulltest} by the green triangles, to be compared to the actual MaxBCG measurement given by the red diamonds. The left-hand panel presents the null test over the full richness range, while the right-hand panel affords an expanded view of the low mass end.  The analysis on the randomised catalog remains consistent with zero (no detection) to within the SZ measurement uncertainty over the entire richness range.  The actual measurements of the MaxBCG clusters, on the other hand, deviate by many $\sigma$ from zero.  We reject the null hypothesis in all bins at high significance.

\begin{figure*}
\centering
\includegraphics[scale=0.5]{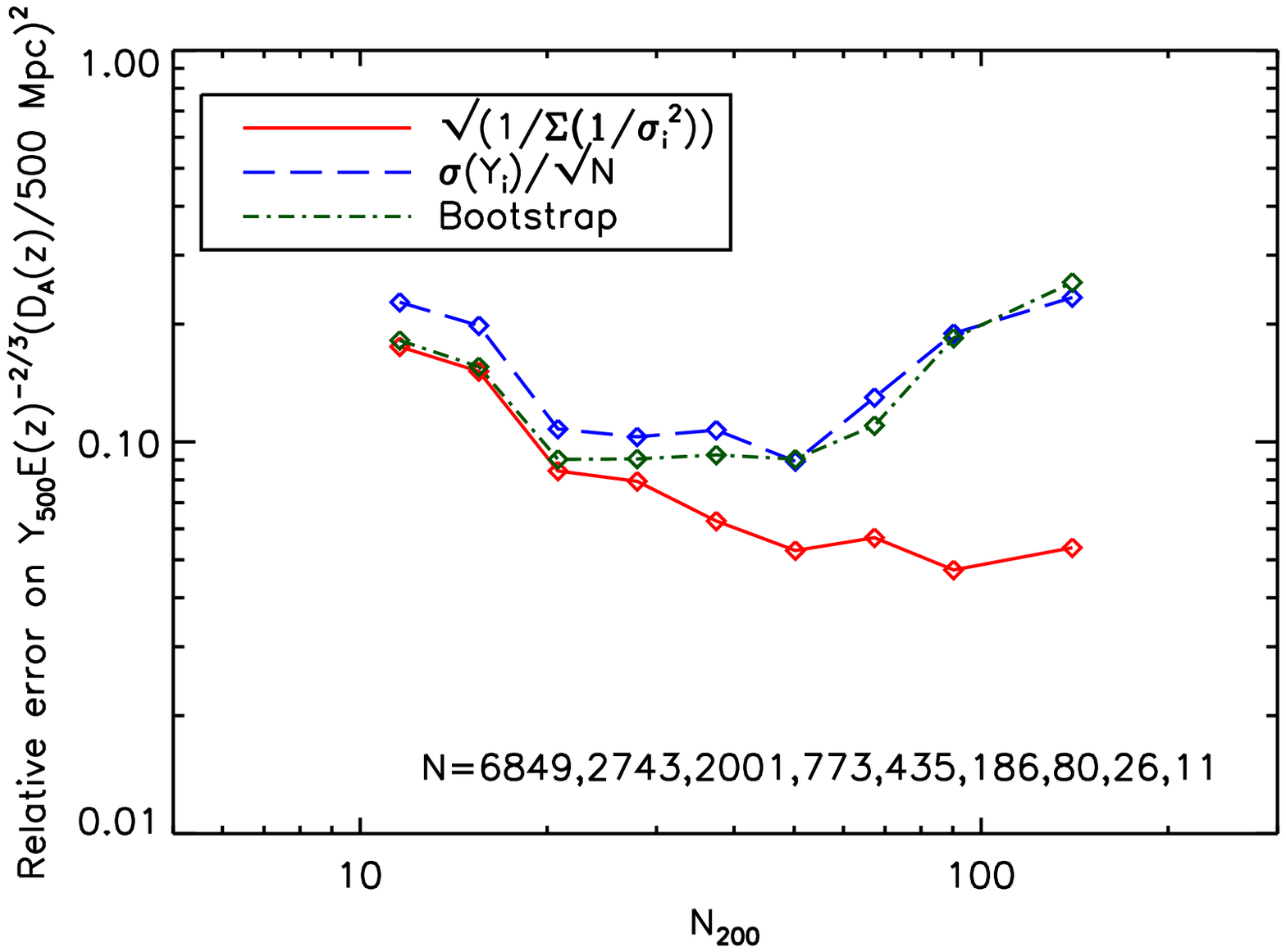}
\includegraphics[scale=0.5]{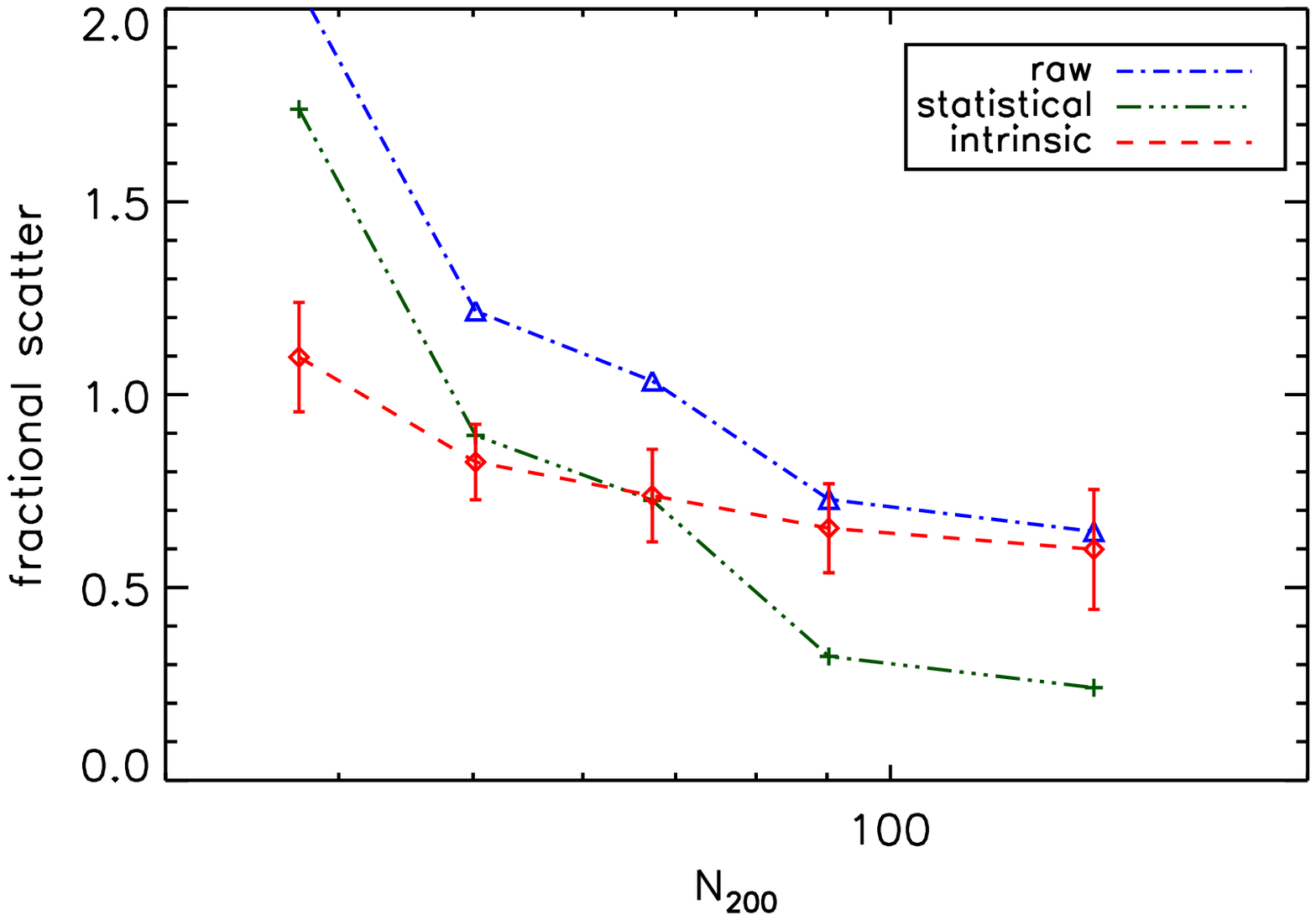}
\caption{Dispersion analysis.  {\em Left-hand panel:\/} relative uncertainty on the mean versus richness.  The relative uncertainty is expressed as a fraction of the bin-average redshift-scaled SZ signal: $\sigma/\Yscaled$.  The lower red curve corresponds to pure measurement uncertainties from the matched filter noise estimations; they are the solid error bars of Fig.~\ref{fig:Ybinned}.  The upper blue curve traces the uncertainty on the mean assuming the points within a richness bin are normally distributed according to the observed in-bin dispersion.  Bootstrap uncertainties are given as the middle green line, found as the dispersion in the mean $\Yscaled$ in each bin calculated over 10,000 bootstrap realisations of the entire MaxBCG catalogue.  The numbers given in the legend indicate the number of objects in each richness bin.  {\em Right-hand panel:\/}  Fractional intrinsic scatter as a function of richness.  The blue dot-dashed line (connecting the blue triangles) shows the raw dispersion in each richness bin, while the green dash-three-dotted line (connecting the green crosses) gives the calculated statistical dispersion from the measurement error on the scaled SZ signal $\Yscaled$.  The red dashed line with error bars is our estimation of the intrinsic scatter as a function of richness.  For this calculation we have eliminated outliers in each bin at $>5\sigma$, with $\sigma = \sigscaled$ for each cluster.  We only calculate the intrinsic scatter at $N_{200}>30$, because at lower richness it becomes difficult to separate the intrinsic dispersion from the scatter due to pure measurement error.}
\label{fig:scatter}
\end{figure*}

Figure \ref{fig:scatter} summarises our analysis of the uncertainty and intrinsic scatter as a function of richness.  In the left panel we show the uncertainty on the mean signal $\Yscaled$ in each bin, expressed as a fraction of $\Yscaled$.  The red solid red line traces the uncertainty on the mean signal due to just the measurement error, i.e., the noise level in the filter.  The blue dashed line gives the uncertainty on the mean assuming that the measurements within a bin are Gaussian distributed about the mean with variance equal to the empirical in-bin variance.  
We show the relative uncertainty calculated from a bootstrap analysis of the entire catalogue as the dot-dashed, green curve.  We perform our full analysis on 10,000 bootstrap realisations from the actual catalogue and use the distribution of the resulting bin averages to find the relative uncertainty.  
The difference between the bootstrap and measurement uncertainties (red line) towards higher richness represents a detection of intrinsic scatter in those bins.  At $N_{200}<30$, this difference is small and any intrinsic scatter is difficult to distinguish from the measurement errors.

In the right-hand panel of Fig.~\ref{fig:scatter} we show our estimate of the intrinsic scatter in the scaling relation as a function of richness for $N_{200}>30$.  This is expressed as a fraction of the mean, $\Yscaled$.  The dot-dashed, blue line traces the empirical, or raw, dispersion around the average signal of each bin.  The three-dot-dashed, green line gives the dispersion corresponding to pure SZ measurement noise.  To find the intrinsic scatter, we use the estimator:
\begin{equation} 
\Sigma^2_{\rm b} = \frac{1}{N_{\rm b}-1}\sum_{i=1}^{N_{\rm b}}\left(\Yscaled(i)-[\Yscaled]_{\rm arith}\right)^2 - \frac{1}{N_{\rm b}}\sum_{i=1}^{N_{\rm b}}\sigscaled^2(i)
\end{equation}
where $[\Yscaled]_{\rm arith}$ is the straight arithmetic mean in the bin.  In the figure we plot $\Sigma_{\rm b}/\langle \Yscaled \rangle_{\rm b}$, with $\langle\Yscaled\rangle_{\rm b}$ being the weighted mean, as above.  For this calculation we clip all outliers at $>5\sigma$, where $\sigma$ is the individual cluster SZ signal error.  The final result, especially at low richness, depends on the chosen clipping threshold.  The scatter is not Gaussian, as the large fractional intrinsic scatter at low richness suggests.  Below $N_{200}\approx 30$, it becomes difficult to draw clear conclusions concerning the scatter, as can be appreciated by the fact that the bootstrap and pure SZ measurement uncertainties begin to overlap in the left-hand panel.  For this reason, we only calculate the intrinsic scatter for the five highest richness bins in the right-hand panel.

In conclusion, we detect a signal down to the lowest mass systems in the MaxBCG catalog with high statistical significance.   {\em This is the central result of our study}.  According to the mass calibration from \citet{johnston2007},  we observe the SZ signal in objects of mass as low as $M_{500}=(4-5)\times 10^{13}\,\Msun$.

\section{Discussion}

Figure~\ref{fig:Ybinned} summarises the central results of our study.  There are two notable aspects: firstly, we detect the SZ signal at high significance over the entire mass range; moreover, simple power laws adequately represent the observed scaling relations.  Secondly, we see a discrepancy in the $\Yscaled-N_{200}$ relation relative to expectations based on X-ray models and either the \cite{johnston2007} or \cite{rozo2009} mass calibrations.

Fitting a power law of the form
\begin{equation}
\label{eq:YNrelation}
\Yscaled = Y_{500}E^{-2/3}(z)\left(\frac{\Dang(z)}{500\,{\rm Mpc}}\right)^2= \YN\left(\frac{N_{200}}{20}\right)^{\aN}
\end{equation}
directly to the individual scaled measurements (e.g., Fig. \ref{fig:Yall}), we obtain the results summarised in Table \ref{tab:fits}.
The \citet{rozo2009} mass calibration assigns a larger mass to the clusters, increasing the filter scale and augmenting the measured SZ signal, which we see as the slightly higher normalisation.  
These fits are plotted as the dashed lines in Fig. \ref{fig:Ybinned}.  The power laws satisfactorily represent the bin-average trends.  The reduced $\chi^2=1.16$ (13,104-2 degrees-of-freedom) in both cases is poor; this reflects the presence of the intrinsic scatter, also evident by the larger uncertainties on the fit from the bootstrap analysis.  

\begin{figure}
\centering
\includegraphics[scale=0.5]{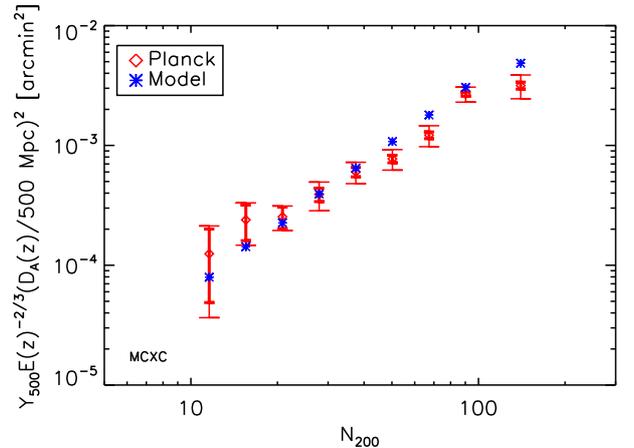}
\caption{The $\Yscaled-N_{200}$ relation for the MCXC X-ray subsample.  Thick lines give the statistical errors, while the thin bars are the bootstrap uncertainties.  We find that the MCXC X-ray subsample matches the model predictions much better than the full sample, which maintains a clear offset relative to the model, as seen in in Fig. \ref{fig:Ybinned}.}
\label{fig:X-raySubSample}
\end{figure}

\begin{table*}[tmb] 
\begingroup 
\newdimen\tblskip \tblskip=5pt
\caption{The SZ signal-richness relation fit to a power law of the form $\Yscaled=\YN\left(N_{200}/20 \right)^\aN$ [see Eq. (\ref{eq:YNrelation})] for the two mass-richness relations.  The power law is fit directly to the individual SZ measurements (e.g., Fig. \ref{fig:Yall}).  The columns labeled ``Statistical'' give the uncertainty on the parameters calculated from the SZ measurement errors alone, while those labeled  ``Bootstrap'' give uncertainties found by fitting the power law to a set of bootstrap samples; the latter better represent the full uncertainty of the fits in the presence of intrinsic scatter.}
\label{tab:fits}
\nointerlineskip
\vskip -3mm
\footnotesize 
\setbox\tablebox=\vbox{ %
\newdimen\digitwidth 
\setbox0=\hbox{\rm 0}
\digitwidth=\wd0
\catcode`*=\active
\def*{\kern\digitwidth}
\newdimen\signwidth
\setbox0=\hbox{+}
\signwidth=\wd0
\catcode`!=\active
\def!{\kern\signwidth}
\halign{ #\hfil \hspace{0.2cm} & \hfil #\hfil \hspace{0.2cm} & \hfil #\hfil \hspace{0.2cm} & \hfil #\hfil \hspace{0.2cm} & \hfil #\hfil \hspace{0.2cm} & \hfil #\hfil  \hspace{0.2cm} & \hfil #\hfil \cr
\noalign{\doubleline}
Mass-Richness Relation & $\YN/(10^{-5}\,{\rm arcmin}^2)$ & Statistical & Bootstrap & $\aN$ & Statistical & Bootstrap \cr
\noalign{\vskip 3pt\hrule\vskip 5pt}
\citet{johnston2007}      & 6.8                                              & $\pm 0.3$         & $\pm 0.4$      & 2.07   & $\pm 0.03$      & $\pm 0.07$       \cr
\citet{rozo2009}             & 7.4                                              & $\pm 0.3$         & $\pm 0.4$      & 2.03   & $\pm 0.03$      & $\pm 0.07$       \cr
\noalign{\vskip 5pt\hrule\vskip 3pt}}}
\endPlancktable 
\endgroup
\end{table*}

The blue stars in Figure~\ref{fig:Ybinned} represent the predictions of a model  based on the $Y_{500}-M_{500}$ relation from \cite{arnaud2010}  and the \cite{johnston2007} (left) or \cite{rozo2009} (right) $M_{500}-N_{200}$ mean scaling relation.  It assumes a self-similar $Y_{500}-M_{500}$ scaling relation (STD case) calibrated on X-ray observations of the REXCESS cluster sample \citep{bohringer2007}.  This calibration is also consistent with WMAP observations \citep{melin2010} and with the \Planck\ analysis \citep{planck2011-5.2a, planck2011-5.2b}.  In each bin we average the model predictions in the same way as the \Planck\ observations: we find the model bin-average redshift-scaled SZ signal as the inverse-error-weighted (pure SZ measurement error) average, assigning each cluster in the bin the same error as the actual observation of that object.  Note that in the observation plane $(\Yscaled, N_{200})$, the model (blue) points change with the mass calibration much more than the measurements.

We see a clear discrepancy between the model and the \Planck\ SZ measurements for both mass calibrations.   In the case of the \citet{johnston2007} mass calibration, the discrepancy manifests as a shift in normalisation that we can characterise by a $25$\% mass shift at given SZ signal: $M\longrightarrow 0.75M$; the slope of the observed relation remains consistent with the self-similar prediction.  The \citet{rozo2009} mass calibration, on the other hand, flattens the mass-richness relation and predicts a shallower power law, as well as a higher normalisation; at $N_{200}=50$ there is a factor of $2$ between the predicted and observed amplitudes.

We now discuss some possible explanations for this discrepancy.  Weak lensing mass estimates are difficult, and as we have seen there is an important difference in the two mass calibrations.  \cite{rozo2009}, building on earlier work by \cite{mandelbaum2008b}, discuss some of the issues when measuring the weak-lensing signal for the MaxBCG catalogue.  However, it seems unlikely that the weak-lensing mass calibration would be in error to the extent needed to explain the discrepancy seen in Figure \ref{fig:Ybinned}.  The discrepancy is in fact larger for the \cite{rozo2009} result, which should be the more robust mass calibration. 

Our model predictions use a series of non-linear, mean relations between observables which in reality have scatter that may also be non-Gaussian.  The largest scatter is expected to be in the mass-richness relation.  If the scatter is large enough, it could bias the predictions.  We have investigated the effect of a 45\% log-normal scatter in mass at fixed richness (e.g., \citet{rozo2009}) and of a Poissonian distributio in richness at fixed mass.  These are realistic expectations for the degree of scatter in the relations.  The effect on the predicted, binned SZ signal is at most 20\%, not enough to explain the factor of two discrepancy we see.

Contamination of the MaxBCG catalogue with a fraction, $f$, of objects that do not contribute an SZ signal (e.g., projection effects in the optical) would bias the measured signal low by about $1-f$.  The level of contamination needed to explain the magnitude of the discrepancy with the
\citet{rozo2009} calibration ($f\approx 0.5$) seems unlikely.  The catalogue is estimated, instead, to be close to 90\% pure for $N_{200}>10$.  Moreover, contamination would also lower the weak-lensing mass calibration by about $1-f$, at given $N_{200}$.  Since the predicted SZ signal scales as $M^{5/3}$, the model SZ signal would drop by an even larger amount than the observed signal.  

To investigate this discrepancy further we analyse, in the same manner, a subsample of the MaxBCG clusters with X-ray data from the MCXC catalogue \citep{piffaretti2010}.  This represents an X-ray detected subsample of the MaxBCG.  The results are given in Fig.~\ref{fig:X-raySubSample}  for the \citet{rozo2009} mass calibration and with our usual notation.  We see that this X-ray subsample, of 189 clusters, matches the model predictions much better.  This argues that, at least for this subsample, the weak-lensing mass calibration is not significantly biased.  The result hints at the presence of two populations: an X-ray under-luminous sample, with a low SZ signal normalisation, seen only in the optically selected catalogue, and an X-ray normal sample matching X-ray model predictions. We emphasize, however, that the selection function of this X-ray subsample is not known, so this interpretation is provisional.  

Splitting the catalogue according to the luminosity of the BCG lends support to the presence of two populations.  In each bin, we divide the catalogue into a BCG-dominated sample, where the fraction of the cluster luminosity contributed by the BCG is larger than the average for that bin, and its complement sample.  The BCG-dominated sample has a notably higher normalisation, closer to the predicted relation, than the complement sample.

We also compare our results to the X-ray results from \citet{rykoff2008b} who stacked ROSAT photons around MaxBCG clusters according to richness.  As with our SZ observations, their individual X-ray fluxes had low signal-to-noise, but they extracted mean luminosities from each image stack \citep{rykoff2008a}.  They report luminosities, $L_{200}$, over the $0.1-2.4\,$keV band and within
$R_{200}$ for each $N_{200}$ richness bin.  Their analysis revealed a  discrepancy between the observed mean luminosities and the  X-ray model predictions, using the \cite{johnston2007} mass calibration.  

\begin{figure}
\centering
\includegraphics[scale=0.5]{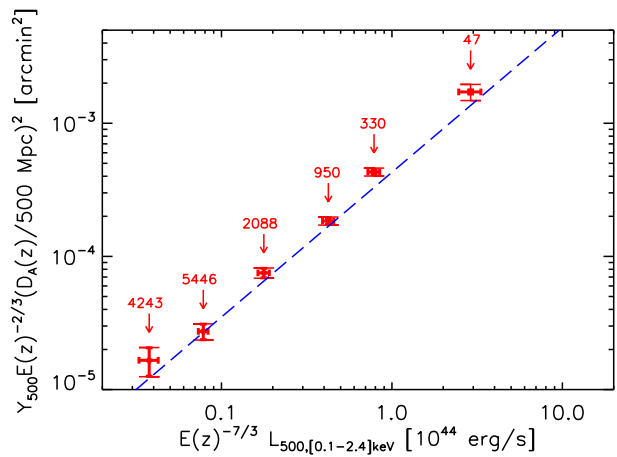}
\caption{Comparison of our bin-average redshift-scaled SZ signal measurements with the mean X-ray luminosities found by \citet{rykoff2008b}.  For this comparison, we re-bin into the same bins as \citet{rykoff2008b} and plot the results as the red diamonds with error bars.  The X-ray luminosities are brought to equivalent $z=0$ values using the self-similar scaling of $E^{-7/3}(z=0.25)$ applied at the quoted $z=0.25$ median redshift.  The dashed blue line shows the predictions of the X-ray model.  Our notation for the error bars follows previous figures.  The numbers in the figure indicate the number of clusters in each bin.
}
\label{fig:rykoff}
\end{figure}

To compare we re-binned into the same richness bins as \cite{rykoff2008a}, calculating the new, bin-average, redshift-scaled $\Yscaled$.  We also convert their luminosities to $L_{500}$ using the X-ray profile adopted in \citet{arnaud2010}; the conversion factor is $0.91$.  In addition, we apply the self-similar redshift luminosity scaling of $E^{-7/3}(z=0.25)$ to bring the \citet{rykoff2008b} measurements to equivalent $z=0$  values from the values at their median redshift, $z=0.25$.  The resulting points are shown in Fig.~\ref{fig:rykoff}.

The model line in the figure is calculated from the $z=0$, X-ray luminosities using the X-ray based scaling laws in \cite{arnaud2010}.   In this plane the model matches the observations well, demonstrating consistency between the SZ and X-ray observations.  Remarkably, the ICM quantities remain in agreement with the model despite the individual discrepancies (SZ and X-ray luminosity) with richness. 

The intrinsic scatter in the scaling relation, given in Figure \ref{fig:scatter}, starts at about 60\% and rises to over 100\% at $N_{200}\approx 30$.  This was calculated by clipping all outliers at $>5\sigma$; the result depends on the choice of clipping threshold, indicative of a non-Gaussian distribution.  This dispersion should be compared to the estimated log-normal scatter in the mass-richness relation of $(45_{-0.18}^{+0.2})$\% found by \cite{rozo2009}.  Assuming that the dispersion in the SZ signal-mass relation is much smaller, we would expect a dispersion of order 75\%, not far from what we find and within the uncertainties.  Such large fractional dispersion implies a non-Gaussian distribution skewed toward high SZ signal values, particularly at low richness.

\section{Conclusions}

We have measured with high significance the mean SZ signal for MaxBCG clusters binned by richness, even the poorest systems.  The observed SZ signal-richness relation, based on 13,104 of the MaxBCG clusters observed by \Planck, is well represented by a power law.  This adds another scaling relation to the list of such relations known to exist among cluster properties and that present important constraints on cluster and galaxy evolution models.  

The observed relation has a significantly lower amplitude than predicted by X-ray models coupled with the mass-richness relation from weak-lensing observations.  The origin of this discrepancy remains unclear.  Bias in the weak-lensing mass measurements and/or a high contamination of the catalogue are potential explanations; another is the presence of a wide range of ICM properties at fixed richness, of which only the more X-ray luminous are readily found in X-ray samples used to establish the X-ray model.  The better agreement of the model with a subsample of the MaxBCG catalogue with X-ray observations supports this conjecture, although the unknown selection function of the X-ray subsample renders it provisional.  Remarkably, the relation between mean SZ signal and mean X-ray luminosity for the entire catalogue does conform to model predictions despite discrepant SZ signal-richness and X-ray luminosity-richness relations; properties of the gas halo appear more stably related than either to richness.

We find large intrinsic scatter in the SZ signal-richness relation, although consistent with the major contribution arising from scatter in the mass-richness relation.  The uncertainties, however, are important.  Such large scatter implies a non-Gaussian distribution of SZ signal at given richness, skewed towards higher signal strengths.  This is consistent with the idea of a wide range of ICM properties at fixed richness, with X-ray detected objects preferentially at the high SZ signal end.

If the lower normalisation of the $\Yscaled-N_{200}$ relation, and by consequence the $\Yscaled-M_{500}$ relation, is confirmed as a physical and representative property of the cluster population, as opposed to selection or other observational effects, then this would have important implications for our understanding of clusters.  Among others, it would imply a lower amplitude for both the diffuse SZ power spectrum and for number counts of SZ-detected clusters.  Predictions for these two quantities depend on the $\Yscaled-M_{500}$ relation.  The amplitude of the SZ power spectrum varies as the square of the normalisation, while the counts depend on it exponentially.  In both instances, this relation represents a significant theoretical uncertainty plaguing models.   

Our study of the SZ signal-richness relation is a step towards reducing this uncertainty, and it presents a new cluster scaling relation as a useful constraint for theories of cluster and galaxy evolution.  Concerning the latter, we find no obvious sign of an abrupt change in the ICM properties of optically selected clusters over a wide range of richness, hence mass, as might be expected from strong feedback models.  Future research with Planck will extend this work to other catalogues and a greater redshift range.
 
\begin{acknowledgements}
The authors from the consortia funded
principally by CNES, CNRS, ASI, NASA, and Danish Natural Research
Council acknowledge the use of the pipeline running infrastructures
Magique3 at Institut d'Astrophysique de Paris (France), CPAC at
Cambridge (UK), and USPDC at IPAC (USA). 
We acknowledge the use of the HEALPix package (\citet{gorski2005}). A description of the Planck Collaboration and a list of its members,
indicating which technical or scientific activities they have been involved in, can
be found at http://www.rssd.esa.int/Planck.

\end{acknowledgements}
\bibliographystyle{aa}
\bibliography{Planck2011-5.2c_bib.bib}

\end{document}